# Pressure-dependent "Insulator-Metal-Insulator" Behavior in Sr-doped La$_3$Ni$_2$O$_7$

Mingyu Xu[1], Shuyuan Huyan[2,3], Haozhe Wang[1], S. L. Bud'ko[2,3], Xinglong Chen[4], Xianglin Ke[5], J. F. Mitchell[4], P. C. Canfield[2,3], Jie Li[6*], Weiwei Xie[1*]

[1]Department of Chemistry, Michigan State University, East Lansing, MI, 48824, USA
[2]Ames National Laboratory, Iowa State University, Ames, IA, 50011, USA
[3]Department of Physics and Astronomy, Iowa State University, Ames, IA 50011, USA
[4]Materials Science Division, Argonne National Laboratory, Lemont, IL, 60439, USA
[5]Department of Physics and Astronomy, Michigan State University, East Lansing, MI, 48824, USA
[6]Department of Geological Sciences, University of Michigan, Ann Arbor, MI, 48109, USA

## Abstract

Recently, superconductivity at high temperatures has been observed in bulk La$_3$Ni$_2$O$_{7-\delta}$ under high pressure. However, the attainment of high-purity La$_3$Ni$_2$O$_{7-\delta}$ single crystals, exhibiting controlled and homogeneous stoichiometry through the post-annealing process in an oxygen-rich floating zone furnace, remains a formidable challenge. Here, we report the crystal structure and physical properties of single crystals of Sr-doped La$_3$Ni$_2$O$_7$ synthesized at high pressure (20 GPa) and high temperature (1400 °C). Through single crystal X-ray diffraction, we showed that high-pressure-synthesized paramagnetic Sr-doped La$_3$Ni$_2$O$_7$ crystallizes in an orthorhombic structure with Ni–O–Ni bond angles of 173.4(2)° out-of-plane and 175.0(2)° and 176.7(2)° in plane. The substitution of Sr alters in band filling and the ratio of Ni$^{2+}$/Ni$^{3+}$ in Sr-doped La$_3$Ni$_2$O$_7$, aligning them with those of "La$_3$Ni$_2$O$_{7.05}$", thereby leading to significant modifications in properties under high pressure relative to the unsubstituted parent phase. At ambient pressure, Sr-doped La$_3$Ni$_2$O$_7$ exhibits insulating properties, and the conductivity increases as pressure goes up to 10 GPa. However, upon further increasing pressure beyond 10.7 GPa, Sr-doped La$_3$Ni$_2$O$_7$ transits back from a metal-like behavior to an insulator. The insulator-metal-insulator trend under high pressure dramatically differs from the behavior of the parent compound La$_3$Ni$_2$O$_{7-\delta}$, despite their similar behavior in the low-pressure regime. These experimental results underscore the considerable challenge in achieving superconductivity in nickelates.

# Introduction

Since the discovery of superconducting cuprates in the last century, nickelates have emerged as potential candidates to harbor this exotic state of matter, primarily owing to the proximity of Ni, a magnetic element, to Cu in the periodic table.[1],[2],[3] A widely recognized explanation for the high-$T_c$ superconductivity involves the hypothesis that it originates from the doping of hole carriers into the Mott insulating state with a half-filled Cu $3d^9$ electronic configuration and $S = 1/2$ spin state.[4],[5] Subsequent to the discovery of superconductivity in cuprates, extensive efforts have been invested in exploring nickel-oxide compounds due to the similarities to cuprates.[6],[7] Infinite-layer nickelates, a well-investigated family, feature $Ni^+$ ($3d^9$) with an electronic configuration akin to $Cu^{2+}$ cations.[8],[9] Earlier investigations reported superconductivity in nickelate thin films, such as Sr-doped $NdNiO_2$ and later $PrNiO_2$, with the $T_c$ values ranging from 9 to 15 K.[10],[11],[12] Notably, superconductivity was confined to the reduced Ruddlesden–Popper phases in thin film form.[13] A breakthrough occurred in May 2023 when superconductivity was reported in the bilayer $La_3Ni_2O_7$ under high pressure.[14] The maximum $T_c$ reached 80 K within the pressure range of 14.0 GPa to 43.5 GPa. Additionally, a potential structural transition was postulated in the high-pressure phase of $La_3Ni_2O_7$ around 14 GPa.[15] At ambient pressure, the Ruddlesden–Popper structure of $La_3Ni_2O_7$ exists two polymorphs (1313 *Cmmm* and 2222 *Cmcm*).[16],[17] The $NiO_6$ octahedra in 2222 *Cmcm* exhibit rotational alignment along the *c*-axis, with the Ni-O-Ni angle deviating from 180° to prevent the formation of the regular square net observed in some high-$T_c$ cuprates.[18],[19],[20] Based on these findings, nickelates emerge as promising compounds for the discovery and study of high-temperature superconductors.

In order to induce a superconducting transition at approximately 80 K, a high pressure of around 14 GPa seems necessary. The manifestation of superconductivity is reported to coincide with a structural transition from *Amam* to *Fmmm*.[14],[21] Given that the structural transition is of the first-order type, quenching the material from high pressure and high temperature may lead to the direct formation of a metastable *Fmmm* phase, enabling the achievement of superconductivity at ambient pressure. Consequently, the high-pressure and high-temperature synthesis emerges as a valuable methodology for investigating and obtaining high-temperature superconductors under ambient pressure conditions.[22],[23] Simultaneously, the 80 K transition temperature provides the possibility for exploring superconductivity through substitution. Utilizing Sr substitution allows for the tuning of band filling and the alteration of the $Ni^{2+}/Ni^{3+}$ ratio, presenting significant opportunities for unraveling the novel superconducting properties of nickelates.

In our pursuit to comprehend the relationship between high-temperature superconductivity in nickelates, band structure, and Ni valence states, we synthesized Sr-doped $La_3Ni_2O_7$ under high pressure (20 GPa) and high temperature (1400 °C). Phase identification and chemical compositions of obtained single crystals of Sr-doped $La_3Ni_2O_7$ were confirmed through single crystal X-ray diffraction and energy dispersive spectroscopy. Post-synthesis, Sr-doped $La_3Ni_2O_7$ crystallizes in the structure of RP bilayer nickelate (*Cmcm*), with an out-of-plane Ni–O–Ni bond angle of 173.4(2)°, while the in-plane Ni–O–Ni bond angles were determined to be 175.0(2)° and 176.7(2)°. The temperature-dependent evolution of the crystal structure was elucidated. As shown in Supporting Information, comprehensive ambient pressure magnetization measurements were performed on a substantial sample quantity. The investigation of resistance and magnetization under high pressure was conducted utilizing single crystals. In the case of Sr-doped $La_3Ni_2O_7$, no superconductivity was detected. On the contrary, the insulator-metal-insulator progression under increasing pressure observed in Sr-doped $La_3Ni_2O_7$ deviates from the documented metal-insulator-metal-superconductivity transition reported in $La_3Ni_2O_7$.

**Experimental Parts**

**Crystal Growth:** Single crystals of Sr-doped $La_3Ni_2O_7$ were obtained in two steps. The first step was the precursor preparation at ambient pressure. Lanthanum (III) oxide (Alfa Aesar REacton, 99.9% powder), nickel (II) oxide (Alfa Aesar, 99% metal based), and strontium peroxide (Thermo Scientific, 12.3% available oxygen) were mixed in a stoichiometric ratio of La:Sr:Ni.[24] The mixed powder was pressed into a pellet using a stainless-steel die press. The pellet was placed in an alumina crucible and sealed in a silica ampoule under vacuum. The ampoule was placed in a furnace, heated to 1000 °C, and kept at the temperature for 10 hours. The precursor for high-pressure synthesis was prepared by grinding the reacted pellet into powder using mortar and pestle. High-pressure synthesis was performed using the Walker-type multi-anvil press at the University of Michigan. The precursor was packed in a platinum capsule, which was placed inside an alumina sleeve, and then inserted into a $LaCrO_3$ heater. The high-pressure assembly used Ceramco octahedral pressure medium and Toshiba-Tungaloy tungsten carbide (WC) anvils.[25] The sample was compressed to 20 GPa at ambient temperature over 18 hours. Then, the sample was heated to 1400 °C and kept for 3 hours. The temperature was quenched to room temperature and 20 GPa pressure was subsequently released. Single crystals were recovered by opening the platinum capsule.

**Crystal Structure Determination and Chemical Composition Confirmation:** A single crystal with dimensions 0.059 × 0.047 × 0.031 $mm^3$ was selected, mounted on a nylon loop with paratone oil, and measured using an XtalLAB Synergy, Dualflex, Hypix single crystal X-ray diffractometer with an Oxford Cryosystems 800 low-temperature device. Data were collected using ω scans with Mo Kα radiation (λ = 0.71073 Å) and Ag Kα radiation (λ = 0.56087 Å, micro-focus sealed X-ray tube, 65 *k*V, 0.67 mA. The total number of runs and images was based on the strategy calculation from the program CrysAlisPro 1.171.43.92a (Rigaku OD, 2023). Data reduction was performed with correction for Lorentz polarization. A numerical absorption correction was applied based on Gaussian integration over a multifaceted crystal model.[26] Empirical absorption correction used spherical harmonics, implemented in SCALE3 ABSPACK scaling algorithm.[27] The structure was solved and refined using the SHELXTL Software Package.[28],[29] The chemical composition of Sr substitution was confirmed by Scanning Electron Microscope energy dispersive spectroscopy

(EDS) quantitative chemical analysis using an EDS detector attached to the 6610LV Scanning Electron Microscope. The composition of a single crystal specimen was measured at several separate positions on each crystal's surface with an acceleration voltage of 20 kV and a working distance of 12 mm.

**Resistance Measurement under High Pressure:** Electrical resistivity measurement by Van der Pauw method was performed in a commercial Diamond Anvil cell (DAC) (Bjscistar[30]) that fits into a Quantum Design Physical Property Measurement System (PPMS). Standard-design-type-Ia diamonds with a culet size of 400 μm were utilized as anvils. A Sr-doped $La_3Ni_2O_7$ single crystal of dimension close to 50 μm × 50 μm × 30 μm was loaded together with a tiny piece of ruby sphere (< 30 μm) into the 250 μm thick, apertured, stainless-steel gasket covered by cubic-boron nitride (c-BN). The sample chamber was about 150 μm in diameter. Platinum foil was used as the electrodes to connect the sample. The NaCl fine powder was used as pressure transmitting medium (PTM). Pressure was determined by measuring the R1 line of the ruby florescent spectra[31].

**Magnetization Measurements under High Pressure:** The DC magnetization measurements under high pressure were performed in a Quantum Design Magnetic Property Measurement System (MPMS) at temperatures down to 5 K. The DAC (easyLab® Mcell Ultra[32]) with a pair of 500 μm culet-size, standard-design-type-Ia diamonds as anvils. The apertured tungsten gasket with 300-μm-diameter hole was used to lock the pressure. The Nujol mineral oil was used as the PTM to provide a more hydrostatic-pressure environment. The applied pressure was measured by the fluorescence line of ruby ball[31]. The background signal of the DAC, with only a piece of ruby sphere (< 30 μm) and PTM in the sample chamber, was measured under 0.5 GPa at 0.1 kOe and 1 kOe, applied magnetic field. Then the DAC was opened and re-closed after loading 6 pieces of tiny Sr-doped $La_3Ni_2O_7$ crystals with dimensions of ~ 50 μm ×50 μm ×30 μm each into the sample chamber. (The ruby sphere was still in the sample chamber). The exact same measurements as previous background measurements were then performed at various pressures. The magnetization of the sample was analyzed by first performing a point-by-point subtraction of long-scan response with/without the sample, and then a dipole fitting of the subtracted long-scan response curve[33].

## Results and Discussion

To determine the crystal structure, single crystal X-ray diffraction measurements were conducted on the samples. The refined crystal structure of Sr-doped $La_3Ni_2O_7$ with $NiO_6$ octahedra was shown in **Fig. 1a**. Our Sr-doped $La_3Ni_2O_7$ crystalizes in the structure of its parent compound, $La_3Ni_2O_7$, with the characteristic Ruddlesden–Popper bilayer stacking, featuring the same pseudo-F-centering problem as what has been well discussed in $La_3Ni_2O_7$.[34],[19],[35] Here, different space groups were examined and finally *Cmcm* was considered as the optimized choice. As for Sr doping, we observe that Sr only goes into La2 site, the one with larger multiplicity and exactly where the rock salt layer is, in the structure. Relaxation of La1 site will not contribute any significant electron density residue for Sr doping. We attribute this as a result that perovskite layers in Ruddlesden–Popper phases are packed more closely and therefore resist external changes to some degree. Oxygen deficiency, always, a topic in the Ruddlesden–Popper bilayer nickelate system, has been checked here. A vacancy was only observed on the shared apical oxygen site (O4) in the perovskite layer. The final single crystal X-ray diffraction refined formula is $La_{2.80(1)}Sr_{0.20(1)}Ni_2O_{6.95(1)}$. The detailed crystal refinement and atomic coordinate information are listed in Table 1 and 2. At room temperature (300 K), the out-of-plane Ni–O–Ni bond angle was determined to be 173.4(2)°, which is larger than that of the *Cmcm* $La_3Ni_2O_7$ structure (168°), whereas the in-plane Ni–O–Ni bonding angles are 175.0(2)° and 176.7(2)° shown in **Fig. 1(b, c)**. To confirm the composition of the crystals, as shown in **Fig. d** and **e**, the secondary electron image of a single crystal was taken. SEM-EDS result, as shown in Supporting Information, of Sr-doped $La_3Ni_2O_7$ is consistent with the stoichiometry determined by single crystal diffraction.

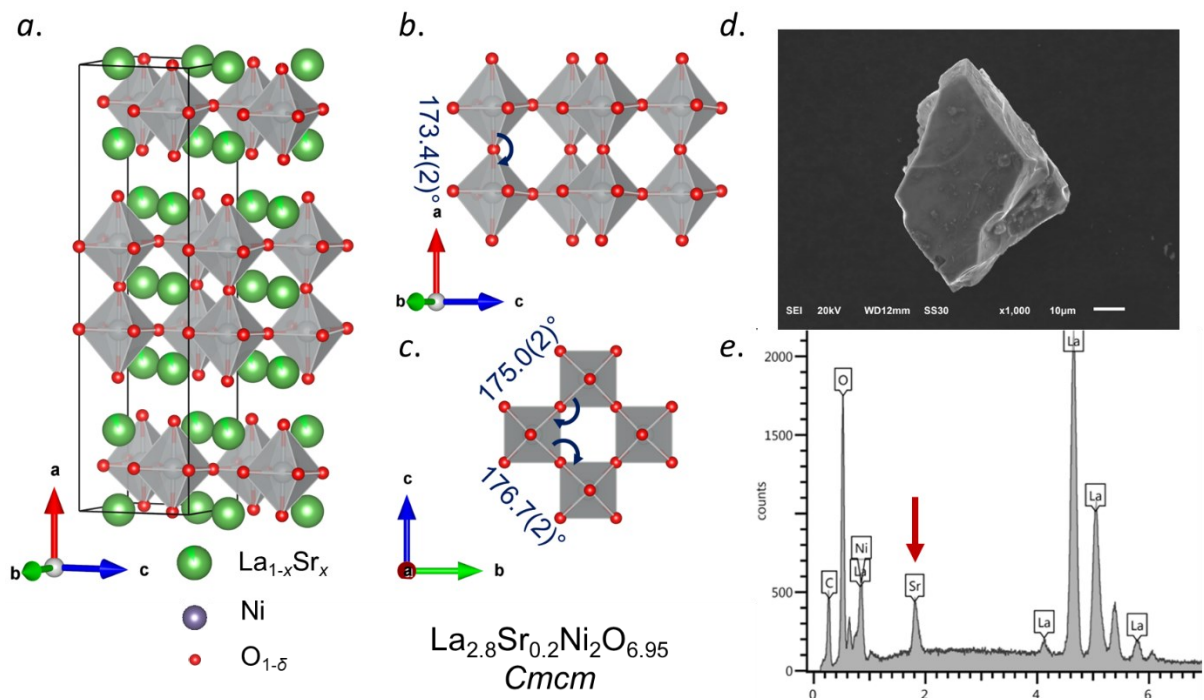

**Fig. 1. Crystal Structure of La$_{2.80(1)}$Sr$_{0.20(1)}$Ni$_2$O$_{6.95(1)}$.** *a.* Unit cell of La$_{2.80(1)}$Sr$_{0.20(1)}$Ni$_2$O$_{6.95(1)}$ (*Cmcm*, #63) with NiO$_6$ octahedra shown. Green, gray, and red indicate La(Sr), Ni, and O atoms, respectively. *b.* View of out-of-plane Ni–O–Ni bond angle of 173.4(2)°. *c.* View of in-plane Ni–O–Ni bond angles of 175.0(2)° and 176.7(2)°. *d.* Secondary electron image of Sr-doped La$_3$Ni$_2$O$_7$ single crystal. *e.* Energy dispersive spectrum with the red arrow pointing at the Lα peak of Sr.

**Table 1.** The crystal structure and refinement of La$_{2.80(1)}$Sr$_{0.20(1)}$Ni$_2$O$_{6.95(1)}$ at 300 K using Mo Kα radiation. Values in parentheses are estimated standard deviation from refinement.

| Chemical Formula | La$_{2.80(1)}$Sr$_{0.20(1)}$Ni$_2$O$_{6.95(1)}$ |
|---|---|
| Formula Weight | 635.89 g/mol |
| Space Group | *Cmcm* |
| Unit Cell dimensions | a = 20.2984(6) Å<br>b = 5.42938(15) Å<br>c = 5.42333(15) Å |
| Volume | 597.69(3) Å$^3$ |
| Z | 4 |
| Density (calculated) | 7.067 g/cm$^3$ |
| Absorption coefficient | 27.524 mm$^{-1}$ |
| F (000) | 1117 |
| 2θ range | 7.96 to 82.50° |
| Reflections collected | 18278 |
| Independent reflections | 1074 [R$_{int}$ = 0.0357] |
| Refinement method | Full-matrix least-squares on F$^2$ |
| Data/restraints/parameters | 1074/0/39 |
| Final *R* indices | R$_1$ (I>2σ(I)) = 0.0142; wR$_2$ (I > 2 σ(I)) = 0.0327<br>R$_1$ (all) = 0.0230; wR$_2$ (all) = 0.0353 |
| Largest diff. peak and hole | +0.868 e$^-$/Å$^3$ and -2.012 e$^-$/Å$^3$ |
| R. M. S. deviation from mean | 0.268 e$^-$/Å$^3$ |
| Goodness-of-fit on F$^2$ | 1.061 |

**Table 2.** Atomic coordinates and equivalent isotropic atomic displacement parameters (Å$^2$) of La$_{2.80(1)}$Sr$_{0.20(1)}$Ni$_2$O$_{6.95(1)}$. ($U_{eq}$ is defined as one-third of the trace of the orthogonalized $U_{ij}$ tensor.) Values in parentheses are estimated standard deviation from refinement.

| Atoms | Wyck. | x | y | z | Occ. | U$_{eq}$ |
|---|---|---|---|---|---|---|
| La1 | 4c | 0 | 0.75074(3) | 1/4 | 1 | 0.00797(4) |
| La2/Sr2 | 8g | 0.32059(2) | 0.24631(2) | 1/4 | 0.898(3)/0.102 | 0.00661(3) |
| Ni | 8g | 0.09682(2) | 0.24930(4) | 1/4 | 1 | 0.00434(5) |
| O1 | 8e | 0.40729(9) | 0 | 0 | 1 | 0.0107(3) |
| O2 | 8e | 0.09970(9) | 0 | 0 | 1 | 0.0121(3) |
| O3 | 8g | 0.20249(9) | 0.2646(3) | 1/4 | 1 | 0.0144(3) |
| O4 | 4c | 0 | 0.2282(5) | 1/4 | 0.954(12) | 0.0156(7) |

**Fig. 2***a* and *b* show the temperature-dependent changes in the lattice parameters for La$_{2.80(1)}$Sr$_{0.20(1)}$Ni$_2$O$_{6.95(1)}$, as determined through single crystal X-ray diffraction. The observed increase in *a* and *c* lattice parameters and volume with increasing temperature aligns with the anticipated behavior of normal thermal expansion. In the observed crystal structure,

the $b$ lattice parameter exhibits minimal variation in comparison to the $a$ or $c$ parameters, particularly below a threshold of 250 K. Notably, as shown in **Fig. 2b**, the out-of-plane Ni-O-Ni angle at room temperature, denoted as $\gamma$, is considerable larger than that in $La_3Ni_2O_7$, while at lower temperature, the angle is smaller, so the corresponding reflection condition has a larger deviation from F-centering. Throughout the single crystal diffraction measurements, multiple samples selected at random from the crystals consistently exhibit the same *Cmcm* structure, affirming the phase as *Cmcm* $La_{2.80(1)}Sr_{0.20(1)}Ni_2O_{6.95(1)}$, and these samples were subsequently employed in the high-pressure measurements.

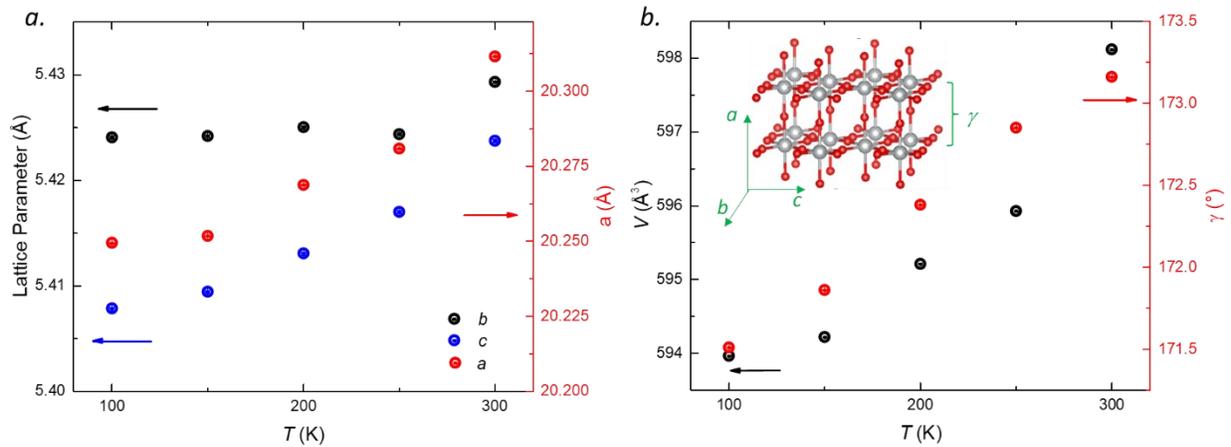

**Fig. 2.** *a*. **Temperature-dependent lattice parameters and Ni-O-Ni bond angle.** Temperature-dependent lattice parameters (*a*: red symbols on the right axis, *b* and *c*: black and blue symbols, respectively, on the left axis) in *Cmcm* Sr-doped $La_3Ni_2O_7$. *b*. temperature-dependent volume of the unit cell (black symbols on the left axis) and the out-of-plane angle of Ni-O-Ni, $\gamma$, as indicated in the inset (red symbols on the right axis).

To investigate the potential existence of superconductivity in Sr-doped $La_3Ni_2O_7$ under high pressure, akin to reports for $La_3Ni_2O_7$, we measured the temperature-dependent electrical resistance up to 19 GPa shown in **Fig. 3**. In **Fig. 3a**, near ambient pressure (1.4 GPa), the resistance exhibits gap-like insulating behavior as a function of temperature. As the pressure increases, the resistance drops and tends to be more metallic like. Upon increasing pressure to 10.7 GPa, the resistance decreases, manifesting metal-like behavior; however, a residual tail persists at this pressure. **Fig. 3b** covers the pressure range from 10.7 GPa to 19 GPa, revealing an increase in resistance with increasing the pressure and a concurrent shift towards a more "insulating" behavior in the temperature-dependent resistance. **Fig. 3** collectively illustrates the absence of discernible phase transitions with changing temperature.

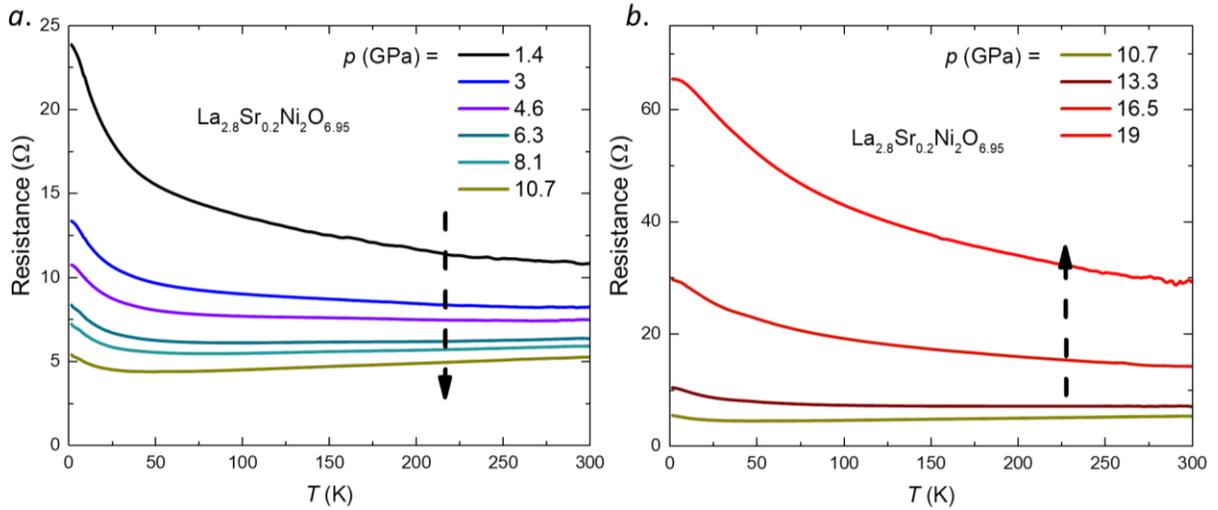

**Fig. 3. Temperature-dependent electrical resistance from 2 K to 300 K at pressures up to 19 GPa.** *a.* Pressure ranges from 1.4 - 10.7 GPa showing the insulator to metal trend as the pressure increases (dashed arrow pointing down). *b.* Pressure ranges from 10.7 - 19 GPa showing the metal to insulator trend as the pressure increases (dashed arrow pointing up).

Although the qualitative low-pressure (< 10.7 GPa) transport behavior of Sr-doped $La_3Ni_2O_7$ resembles that of the parent phase, [14],[15] beyond 10.7 GPa the two behave quite differently, as shown by the normalized resistance (R/R(300K)) plotted in **Fig. 4**. Since the lowest pressure of two compounds is different, to compare the tendency of pressure-dependent resistance of Sr-doped and parent compounds easily, the resistance at 300 K is normalized to the first low pressure data and forces the first data to lie on the intertrapolated trend of parent compound. In $La_3Ni_2O_7$, metallic character increases, ultimately leading to superconductivity. In Sr-doped $La_3Ni_2O_7$, a very similar behavior is observed before the expected phase transition; however, resistance starts to rise above 10.7 GPa. Notably, no superconducting transition was detected above 1.8 K in high-pressure measurements of Sr-doped $La_3Ni_2O_7$.

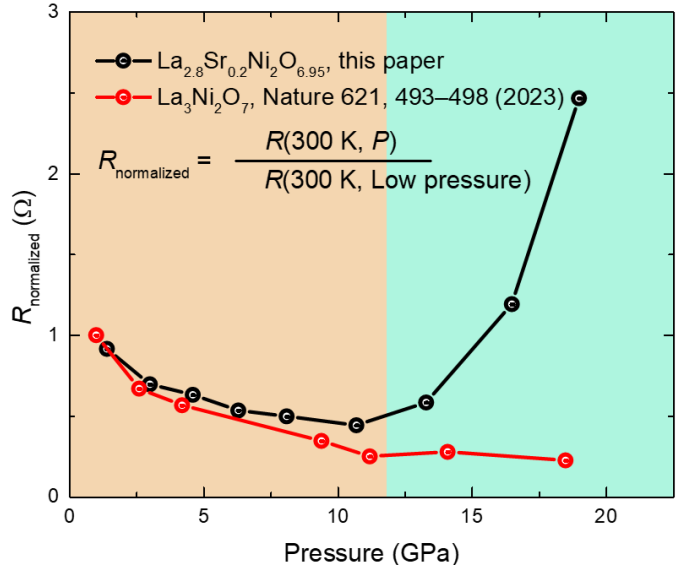

**Fig. 4. Comparison of pressure dependent normalized resistance at 300 K**. The resistance at 300 K is normalized with the first low pressure data by forcing the first data point to line on the extrapolated trend. The data of this paper is given in black, and the data from the reference is given in red. The color edge between pink and green indicates the start of structure transition and superconducting transition in $La_3Ni_2O_7$.

The evaluation of pressure-dependent magnetization was conducted on a single crystal of Sr-doped $La_3Ni_2O_7$, as illustrated in **Fig. 5**, utilizing an externally applied magnetic field of 0.1 kOe for **Fig. 5*a*** and 1 kOe for **Fig. 5*b***. Measurements were systematically carried out under zero-field-cooling-warming (ZFCW) and field-cooling-cooling (FCC) modes, revealing no irreversibility. Over the temperature ranging from 10 K to 200 K and the pressure range extending from 1.6 GPa to 27.1 GPa, the magnetic moment exhibits temperature-independent magnetization. No discernible magnetic transition (with $10^{-6}$ emu resolution), including any indicative of superconductivity, was detected up to the maximum applied pressure of 27.1 GPa.

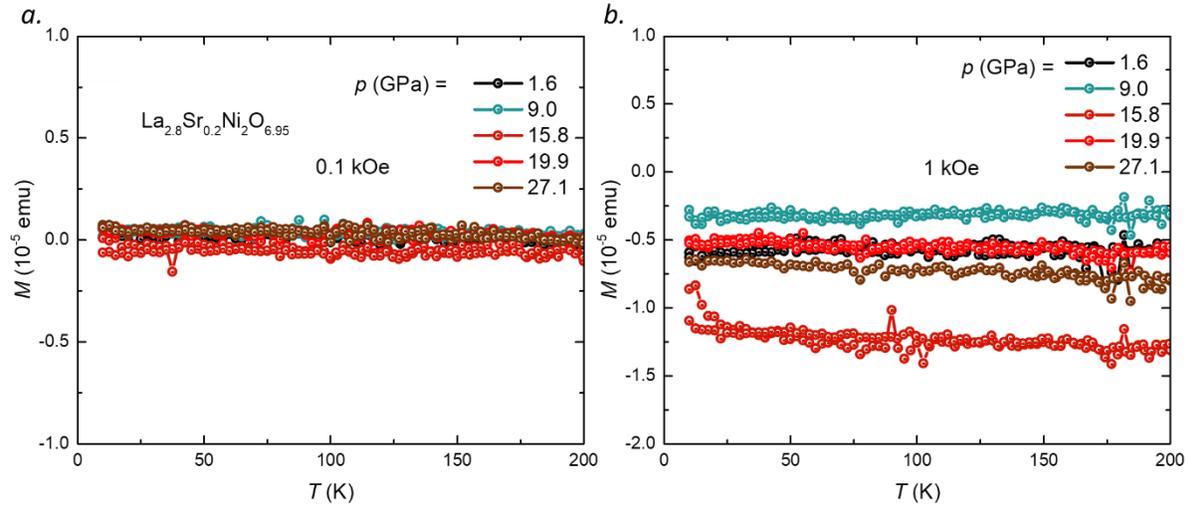

**Fig. 5. Pressure dependence of magnetization of Sr-doped La$_3$Ni$_2$O$_7$ single crystal as a function of temperature (10-200 K) under 0.1 kOe (*a*) and 1 kOe (*b*) magnetic field.** Different colors code pressures in GPa. A background has been subtracted (see text).

## Conclusion

The synthesis undertaken at high pressure and high temperature conditions (20 GPa and 1400 °C) aimed to stabilize the superconducting phase previously observed under high pressure (above 14 GPa). In-depth characterization through single-crystal X-ray diffraction and SEM-EDS was employed to elucidate the crystal structure and phase details. The investigation revealed that the Sr-doped La$_3$Ni$_2$O$_7$ obtained through this synthesis maintains the ambient pressure structure, adhering to the space group of *Cmcm*. The outcomes of high-pressure electrical measurements and magnetization studies reveal no evidence of any superconducting transition in Sr-doped La$_3$Ni$_2$O$_7$ up to 27 GPa. Intriguingly, under high pressure, Sr-doped La$_3$Ni$_2$O$_7$ displays a distinct trend in resistance, transitioning from insulating to metallic and back to insulating. This behavior stands in contrast to the observations in La$_3$Ni$_2$O$_7$. The introduction of Sr substitution in the crystalline lattice is anticipated to introduce hole doping and alter the lattice parameters, thereby influencing band filling and band width within the material. Concurrently, the application of external pressure also exerts an influence on the lattice parameters and is expected to affect the band width, implying that the variations in band width resulting from Sr substitution can potentially be mitigated through the application of pressure. Notably, the observed similarity in electrical resistance behavior under a pressure of 10.7 GPa implies that the lattice parameter modifications caused by Sr substitution are effectively nullified by the increasing pressure. Consequently, it can be inferred that the distinction between Sr-doped samples and their parent compound $La_3Ni_2O_7$ of transiting from metallic and back to insulating after 10.7 GPa. These alterations apparently not only suppress the claimed 80 K superconductivity but also induce significant modifications in the band structure, transitioning from a metallic to an insulating state post 10.7 GPa. Collectively, these findings strongly underscore the fragility of superconductivity in $La_3Ni_2O_7$, emphasizing the inherent challenges associated with its stabilization and realization.

## Acknowledgement

The work at Weiwei Xie Lab was supported by the U.S.DOE-BES under Contract DE-SC0023648. X.K. acknowledges the financial support by the U.S. Department of Energy, Office of Science, Office of Basic energy Sciences, Materials Science and Engineering Division under DE-SC0019259. Work at University of Michigan was supported by NSF grant EAR2310830. Work in Ames was supported by the U.S. Department of Energy, Office of Science, Basic Energy Sciences, Materials Sciences and Engineering Division under Contract No. DE-AC02-07CH11358. Work at Argonne National Laboratory was supported by the Department of Energy, Office of Science, Basic Energy Sciences, Materials Sciences and Engineering Division.

## Reference

[1] J. G. Bednorz, K. A. Müller, *Z. Für Phys. B Condens. Matter* **1986**, *64*, 189.
[2] W. E. Pickett, *Nat. Rev. Phys.* **2021**, *3*, 7.
[3] Y. Nomura, R. Arita, *Rep. Prog. Phys.* **2022**, *85*, 052501.
[4] P. A. Lee, N. Nagaosa, X.-G. Wen, *Rev. Mod. Phys.* **2006**, *78*, 17.
[5] P. Phillips, *Rev. Mod. Phys.* **2010**, *82*, 1719.
[6] B. Keimer, S. A. Kivelson, M. R. Norman, S. Uchida, J. Zaanen, *Nature* **2015**, *518*, 179.
[7] J. Zhang, X. Tao, *CrystEngComm* **2021**, *23*, 3249.
[8] B. Y. Wang, K. Lee, B. H. Goodge, *Annu. Rev. Condens. Matter Phys.* **2024**, *15*.
[9] X. Wu, D. Di Sante, T. Schwemmer, W. Hanke, H. Y. Hwang, S. Raghu, R. Thomale, *Phys.*

*Rev. B* **2020**, *101*, 060504.

[10] D. Li, K. Lee, B. Y. Wang, M. Osada, S. Crossley, H. R. Lee, Y. Cui, Y. Hikita, H. Y. Hwang, *Nature* **2019**, *572*, 624.

[11] G. A. Pan, D. Ferenc Segedin, H. LaBollita, Q. Song, E. M. Nica, B. H. Goodge, A. T. Pierce, S. Doyle, S. Novakov, D. Córdova Carrizales, A. T. N'Diaye, P. Shafer, H. Paik, J. T. Heron, J. A. Mason, A. Yacoby, L. F. Kourkoutis, O. Erten, C. M. Brooks, A. S. Botana, J. A. Mundy, *Nat. Mater.* **2022**, *21*, 160.

[12] H. Lu, M. Rossi, A. Nag, M. Osada, D. F. Li, K. Lee, B. Y. Wang, M. Garcia-Fernandez, S. Agrestini, Z. X. Shen, E. M. Been, B. Moritz, T. P. Devereaux, J. Zaanen, H. Y. Hwang, K.-J. Zhou, W. S. Lee, *Science* **2021**, *373*, 213.

[13] M. Osada, B. Y. Wang, B. H. Goodge, K. Lee, H. Yoon, K. Sakuma, D. Li, M. Miura, L. F. Kourkoutis, H. Y. Hwang, *Nano Lett.* **2020**, *20*, 5735.

[14] H. Sun, M. Huo, X. Hu, J. Li, Z. Liu, Y. Han, L. Tang, Z. Mao, P. Yang, B. Wang, J. Cheng, D.-X. Yao, G.-M. Zhang, M. Wang, *Nature* **2023**, *621*, 493.

[15] Y. Zhou, J. Guo, S. Cai, H. Sun, P. Wang, J. Zhao, J. Han, X. Chen, Q. Wu, Y. Ding, M. Wang, T. Xiang, H. Mao, L. Sun, "Evidence of filamentary superconductivity in pressurized $La_3Ni_2O_7$ single crystals," can be found under https://arxiv.org/abs/2311.12361v1, **2023**.

[16] X. Chen, J. Zhang, A. S. Thind, S. Sharma, H. LaBollita, G. Peterson, H. Zheng, D. Phelan, A. S. Botana, R. F. Klie, J. F. Mitchell, **2023**, DOI 10.48550/arXiv.2312.06081.

[17] H. Wang, L. Chen, A. Rutherford, H. Zhou, W. Xie, **2023**, DOI 10.48550/arXiv.2312.09200.

[18] J. Zhang, A. S. Botana, J. W. Freeland, D. Phelan, H. Zheng, V. Pardo, M. R. Norman, J. F. Mitchell, *Nat. Phys.* **2017**, *13*, 864.

[19] Z. Zhang, M. Greenblatt, J. B. Goodenough, *J. Solid State Chem.* **1994**, *108*, 402.

[20] J. Song, D. Ning, B. Boukamp, J.-M. Bassat, H. J. M. Bouwmeester, *J. Mater. Chem. A* **2020**, *8*, 22206.

[21] P. Puphal, P. Reiss, N. Enderlein, Y.-M. Wu, G. Khaliullin, V. Sundaramurthy, T. Priessnitz, M. Knauft, L. Richter, M. Isobe, P. A. van Aken, H. Takagi, B. Keimer, Y. E. Suyolcu, B. Wehinger, P. Hansmann, M. Hepting, **2023**, DOI 10.48550/arXiv.2312.07341.

[22] J. V. Badding, *Annu. Rev. Mater. Sci.* **1998**, *28*, 631.

[23] V. V. Brazhkin, *High Press. Res.* **2007**, *27*, 333.

[24] J. Zhang, X. Tao, *CrystEngComm* **2021**, *23*, 3249.

[25] K. D. Leinenweber, J. A. Tyburczy, T. G. Sharp, E. Soignard, T. Diedrich, W. B. Petuskey, Y. Wang, J. L. Mosenfelder, *Am. Mineral.* **2012**, *97*, 353.

[26] S. Parkin, B. Moezzi, H. Hope, *J. Appl. Crystallogr.* **1995**, *28*, 53.

[27] N. Walker, D. Stuart, *Acta Crystallogr. A* **1983**, *39*, 158.

[28] G. M. Sheldrick, *Acta Crystallogr. Sect. C Struct. Chem.* **2015**, *71*, 3.

[29] G. M. Sheldrick, *Acta Crystallogr. Sect. Found. Adv.* **2015**, *71*, 3.

[30] "23mm cell," can be found under http://www.bjscistar.com/page169?product_id=127, **n.d.**

[31] G. Shen, Y. Wang, A. Dewaele, C. Wu, D. E. Fratanduono, J. Eggert, S. Klotz, K. F. Dziubek, P. Loubeyre, O. V. Fat'yanov, P. D. Asimow, T. Mashimo, R. M. M. Wentzcovitch, *High Press. Res.* **2020**, *40*, 299.

[32] "easyLab® Mcell Ultra," **n.d.**

[33] R. Cabassi, F. Bolzoni, F. Casoli, *Meas. Sci. Technol.* **2010**, *21*, 035701.

[34] C. D. Ling, D. N. Argyriou, G. Wu, J. J. Neumeier, *J. Solid State Chem.* **2000**, *152*, 517.

[35] V. I. Voronin, I. F. Berger, V. A. Cherepanov, L. Ya. Gavrilova, A. N. Petrov, A. I. Ancharov, B. P. Tolochko, S. G. Nikitenko, *Nucl. Instrum. Methods Phys. Res. Sect. Accel. Spectrometers Detect. Assoc. Equip.* **2001**, *470*, 202.